\documentclass[aps,preprint,showpacs]{revtex4}
\usepackage{graphicx}

\begin{document}

\title{ pre-saddle neutron multiplicity for  fission reactions induced
by  heavy ions and light particles}

\author{S. Soheyli\footnote{Corresponding author: s.soheyli@basu.ac.ir}, M. K. Khalili}
\affiliation{Bu-Ali Sina University, Department of Physics, Hamedan,
 Iran}
\begin{abstract}

Pre-saddle neutron multiplicity has been calculated for several
 fission reactions  induced by  heavy ions and light particles.
Experimentally, it is impossible to determine the contribution of
neutrons being emitted before the saddle point and those emitted
between the saddle and the scission points. Determination of the
pre-saddle neutron multiplicity in our research is based on the
comparison between the experimental anisotropies  and those
predicted by the standard saddle-point statistical model. Analysis
of the results shows that the pre-saddle neutron multiplicity
depends on the fission barrier height and stability of the compound
nucleus. In heavy ion induced fission, the number of pre-saddle
neutrons decreases with increasing the excitation energy of the
compound nucleus. A main cause of this behavior is due to a
reduction in the ground state-to-saddle point transition time with
increasing the excitation energy of the compound nucleus. Whereas in
induced fission by light particles, the number of pre-saddle
neutrons increases with increasing  the excitation energy of the
compound nucleus.
\end{abstract}

\pacs{ 25.70.Jj, 25.85.Ge} \maketitle
\newpage
\section{Introduction}
In the last two decades, much theoretical attention has directed
towards understanding the dynamics  of  fission. According to
reports, measuring the number of neutrons emitted during fission
most likely  gives information on the timescale of fission as well
as on the nuclear dynamics.  The transition state model of fission,
based on appropriate level densities, predicts the widths (and thus
lifetimes) of fission and neutron emission. This model is also
suitable for determining the pre-fission neutron multiplicity if the
calculated lifetimes are long compared to the dynamically
constrained fission lifetime.  Several groups have invested an
extensive effort in measuring the number of emitted neutrons
associated with  fission reactions induced by heavy ions [1-17]. The
measurement of emitted neutrons  is usually limited to the
measurement of pre-scission neutron multiplicity, post-scission
neutron multiplicity, and therefore total neutron multiplicity.
 These measurements show that the transition state model of fission
leads to an underestimation of the number of measured pre-scission
neutrons emitted in heavy ion induced fission at high excitation
energies. This discrepancy can be related to the viscosity of the
hot nucleus [18]. Hence, the fission lifetime of a hot nucleus is
substantially longer than that determined by  statistical model of
Bohr and Wheeler [19]. As a result, it is natural to expect that a
dissipative dynamical model would provide an appropriate description
of nuclear fission at high excitation energies [20].

  Pre-scission neutrons $\nu_{pre}$, can be emitted between the ground state of the compound
nucleus and the saddle point (pre-saddle neutrons) $\nu_{gs}$, or
between the saddle and the scission points (saddle-to-scission
neutrons) $\nu_{ss}$. The number of pre-saddle neutrons as well as
the number of saddle-to-scission neutrons can be determined by a
combined dynamical statistical  model ( CDSM ) [21-23]. The
contributions $\nu_{gs}$ and $\nu_{ss}$ to the pre-scission neutron
multiplicity are also estimated by a stochastic approach based on
three-dimensional Langevin equation [24]. Recently, a more accurate
four-dimensional Langevin model as an extension of the
three-dimensional Langevin model by adding the fourth collective
coordinate ( the projection of the total spin about the symmetry
axis of the fissioning nucleus ) is used to calculate the
pre-scission neutron multiplicity [25].

 A common assumption in the calculation of  the angular anisotropy  of fission fragments using by
 the transition state model is that all pre-scission neutrons are emitted prior to reaching the
saddle-point, since it is not straightforward to separate
experimentally the contribution of neutrons being emitted before the
saddle-point and those emitted between the saddle and  the scission
points [26-35]. It is well known that the standard saddle-point
statistical model (SSPSM) has become the standard theory of fission
fragment angular distributions and received great success since it
was proposed. The effect of neutron evaporation prior to reaching
the saddle-point is to reduce the temperature of the fissioning
nucleus which in turn increases the fission fragment anisotropy
prediction by using this model.  Only $\nu_{gs}$ has an influence
over the prediction of angular anisotropy by using SSPSM.  The upper
limit of the angular anisotropy of fission fragments, based on the
prediction of SSPSM is determined  by assuming that all  the
pre-scission neutrons are emitted before the saddle-point.

  The  pre-saddle neutrons  as a crucial quantity in
determining the angular anisotropy of fission fragments by using
SSPSM plays a main role, although any precise method to determine it
has not been introduced. In this article, we calculate the number of
pre-saddle neutrons by a novel method. In this method, the values of
$\nu_{gs}$ for several induced fission reactions  by light particles
and heavy ions are determined by the fission fragments angular
distribution method. This method is based on comparison between the
experimental anisotropies and those predicted by the standard
saddle-point statistical model. This method is limited to the
calculation of pre-saddle neutrons in induced fission in which the
angular anisotropy of fission fragments has a normal behavior, i.e.,
it is observed a good agreement between the angular anisotropy of
fission fragments and that predicted by the SSPSM.

 In order to make the present paper self-contained,  we present in Sec.
II,  an brief description  of the standard saddle-point statistical
model as well as the calculating method of the pre-saddle neutron
multiplicity on the basis of the SSPSM in detail. Section III is
devoted to the results obtained in this study. Finally, the
concluding remarks are given in Sec. IV.

\section{Method of Calculations }
\subsection{Standard saddle-point Statistical Model}

The  standard transition-state model has been used to analyze the
angular anisotropy of fission fragments in fission. In the
transition-state model, the equilibrium distribution over the K
degree of freedom (the projection of total angular momentum of the
compound nucleus (\textrm{I}) on the symmetry axis of the fissioning
nucleus) is assumed to be established at the transition state. Two
versions of the transition-state models based on assumptions on the
position of the transition state: standard saddle-point statistical
model (SSPSM), and scission-point statistical model (SPSM)  can be
used for the prediction of fission fragment angular distributions.
The basic assumption of the  SSPSM is that fission proceeds along
the symmetry axis of a deformed compound nucleus, and that the
distribution of K   is frozen from the saddle point to the scission
point. In this model, the fission fragment angular distribution
$W(\theta)$ for the fission of spin zero nuclei is given by the
following expression~\cite{Vandenbosch1:1973}

\begin{equation}
W(\theta)\propto\sum_{I=0}^{\infty}\frac{(2I+1)^2T_{I}\exp[-p\sin^2{\theta}]J_{0}[-\emph{i}p\sin^2
{\theta}]}{\textrm{erf}[\sqrt{2p}]}.
\end{equation}
Where  ${T_I}$, and ${J_0}$ are the transmission coefficient for
fission,  and the zeroth-order Bessel function,
$p=(I+\frac{1}{2})^2/(4K_{\circ}^{2})$, and  the variance of the
equilibrium K distribution ($K_{\circ}$) is
\begin{equation}
K_{\circ}^{2}=\frac{\Im_{eff}T}{\hbar^{2}},
\end{equation}

here ${\Im_{eff}}$ and $T$ are the effective moment of inertia and
the nuclear temperature of the compound nucleus at the saddle point,
respectively.

  The angular anisotropy of  fission fragments is
defined as
\begin{equation}
A=\frac{W(0^{\circ})}{W(90^{\circ})}.
\end{equation}

  The nuclear temperature of the compound nucleus at the saddle point
is given by

\begin{equation}
T=\sqrt{\frac{E_{ex}}{\emph{a}}},
\end{equation}
where $E_{ex}$ is the excitation energy of the fissioning system and
$\emph{\textrm{a}}$ is the nuclear level density parameter at the
saddle point. $E_{ex}$ can be expressed by the following relation
\begin{equation}
E_{ex}=E_{c.m.}+Q-B_{f}(I)-E_{R}(I)-{\nu_{gs}}E_{n}.
\end{equation}

In this equation, $E_{c.m.}$, $Q$, $B_{f}(I)$, $E_{R}(I)$,
$\nu_{gs}$, and $E_{n}$ represent the center-of-mass energy of the
projectile, the $Q$ value, the spin dependent fission barrier
height, the spin dependent rotational energy of the compound
nucleus, the number of pre-saddle neutrons, and the average
excitation energy lost due to evaporation of one neutron from the
compound nucleus prior to the system reaching to the saddle point,
respectively. In the case of $p\gg1$, the angular anisotropy of
fission fragments  by using Eq. (1)  is given by the following
approximate relation
\begin{equation}
A\approx1+\frac{<I^{2}>}{4K_{\circ}^{2}}.
\end{equation}

   The prediction of angular anisotropy of fission fragments by using
the SSPSM is valid only under restrictive assumptions. At high
angular momentum, or at high fissility, the rotating liquid drop
model (RLDM) predicts that the fission barrier height($B_{f}(I)$)
vanishes even for a spherical nucleus, which leads to
$K_{\circ}^{2}\rightarrow\infty$. Subsequently, the distribution of
K is uniform and hence the prediction of the SSPSM for the fission
fragments angular anisotropy is nearly uniform by using Eq. (1).
This predicted tendency toward isotropy for fission fragments at
high angular momentum is not seen in the experiments. This
discrepancy is taken as a clear indication that the width of the K
distribution is not determined at the predicted spherical saddle
point shape, but at a point where nucleus is more deformed.
Therefore, it has been proposed that the standard saddle-point
statistical model breaks down at high spin and/or large values of
$\frac{Z^{2}}{A}$ of the compound nucleus (CN), and the angular
distribution of fission fragments is governed  by an effective
transition state different from saddle point transition state.

\subsection{Pre-saddle Neutron multiplicity }

 It is clear that because of the hindrance to fission, a large number
of particles more that those predicted by the statistical model are
emitted from the fissioning system. In heavy ion fusion reactions,
due to the formation of a heavy compound nucleus, the competition
between neutron emission and fission describes the decay
possibilities rather well. During the collective motion to the
scission point, neutrons will be evaporated if energetically
possible, and would be experimentally as pre-fission, or more
correctly, pre-scission neutrons. A longer saddle to scission time
due to the viscosity effect, will result in a higher pre-scission
neutron multiplicity~\cite{Hinde3:1989}. The calculation of
pre-saddle neutrons in heavy ion induced reactions based on the
comparison between the experimental data of angular anisotropy and
those predicted  by the SSPSM depends on the kinetic energy and the
binding energy of evaporated neutron from the compound nucleus prior
to the system reaching to the saddle point. The energy spectrum of
evaporated neutrons is usually given by the following form (an
evaporation spectrum)~\cite{Tsang:1983}
\begin{equation}
\frac{dN}{dE}=CE\exp(-\frac{E}{T}).
\end{equation}
Hence, the average kinetic energy of the emitted neutron,
$\overline{E}_{K}$  is given by
\begin{equation}
\overline{E}_{K}=2T.
\end{equation}
The average excitation energy lost due to evaporation of one neutron
from the compound nucleus prior to the system reaching to the saddle
point is given by
\begin{equation}
E_n=B_n+2T,
\end{equation}
where, $B_n$ denotes the average neutron separation energy.

In this work, the average energy  lost  by an emitted neutron over
the energy range of the projectile is calculated by Eq. (9), for
heavy ion induced fission reactions, as well as for induced fissions
by light projectiles. The level density parameter,
 $\emph{a}$~ is taken $\frac{A_{C. N.}}{8}$ ( Considering the level density parameter
 as   $\frac{A_{C. N.}}{10}$, rather than $\frac{A_{C. N.}}{8}$,
the number of pre-saddle neutrons varies at most by 10$\%$). Hence,
number of pre-saddle neutrons  is not sensitive to the level density
parameter selected in the calculation.  $\Im_{eff}$, $B_{f}(I)$, and
$E_{R}(I)$ are accounted by the use of rotating finite range model
(RFRM)~\cite{Sierk:1986}, while $<I^{2}>$ quantities  are calculated
by several models [39-44]. In the following sections, the
determination of the number of  pre-saddle neutrons, $\nu_{gs}$  for
these systems is based on the comparison between the experimental
data of  angular anisotropies and those predicted by the SSPSM. In
the present work, it is determined pre-saddle neutron multiplicities
for several systems undergoing heavy ion induced fission  in which
fission fragments angular anisotropies have a normal behavior as
well as those systems undergoing light particle induced fission.  In
order to determine number of pre-saddle neutrons in heavy ion
reactions with anomalous angular anisotropies, it is necessary to
predict the average contribution of non compound nucleus fission
events ~\cite{Soheyli:2012}.

\section{Results and discussion}

The calculated multiplicities  of pre-saddle neutrons as a function
of $E_{ex}$  for the two $^{16}
\textrm{O}+^{209}\textrm{Bi}\rightarrow^{225}$ \textrm{Pa} and
$^{19} \textrm{F}+^{208}\textrm{Pb}\rightarrow^{227}\textrm{Pa}$
reaction systems  leading to Protactinium isotopes,  are shown in
Fig. 1(a). For above studied systems, the experimental data of
angular anisotropy are taken from literature [46, 47]. As
illustrated in the figure, the number of pre-saddle neutrons
decreases with increasing the excitation energy of the compound
nucleus. This behavior  is due to the fact that the fission barrier
height (and thus the ground state-to-saddle point transition time )
decreases with increasing the excitation energy of the compound
nucleus, which can be lead to that $\nu_{gs}$ decreases with
$E_{ex}$.  In this figure, the general trend of the number of
pre-saddle neutrons as a function of the excitation energy of the
compound nucleus is represented by a line using the method of least
squares. Fig. 1(b), shows a similar case for the
$^{16}\textrm{O}+^{208}\textrm{Pb}\rightarrow^{224}\textrm{Th}$
reaction system. For this system, the experimental data of angular
anisotropy are  taken from literature [8]. Multiplicities of
pre-saddle neutrons  for the two $^{11}
\textrm{B}+^{237}\textrm{Np}$ and $^{16}
\textrm{O}+^{232}\textrm{Th}$ reaction systems, both populating the
same compound nucleus $^{248}\textrm{Cf}$ are also shown in Fig.
1(c). For these two systems, the experimental data of $<\textrm{A}>$
are taken from literature [56-58]. It is interesting to note that
for these two systems, as well as for the two $^{16}
\textrm{O}+^{209}\textrm{Bi}\rightarrow^{225}$ \textrm{Pa} and
$^{19} \textrm{F}+^{208}\textrm{Pb}\rightarrow^{227}\textrm{Pa}$
reaction systems as shown in Fig. 1(a),  the  number of pre-saddle
neutrons at any given excitation energy appears to be nearly equal.
As a result, the multiplicities of pre-saddle neutrons for heavy ion
fusion reactions populating the same compound nucleus are nearly
independent of  the entrance channel asymmetry and depend  on the
mass number of the compound nucleus.

\begin{figure}[h]
\centering
\includegraphics[scale=0.8]{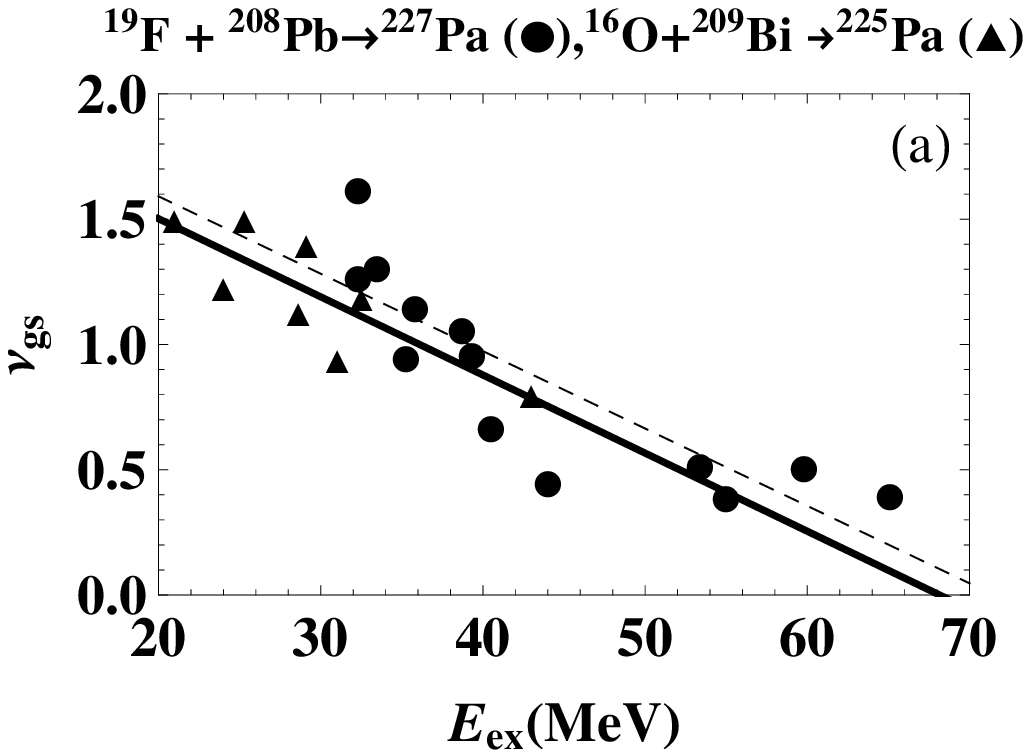}\includegraphics[scale=0.8]{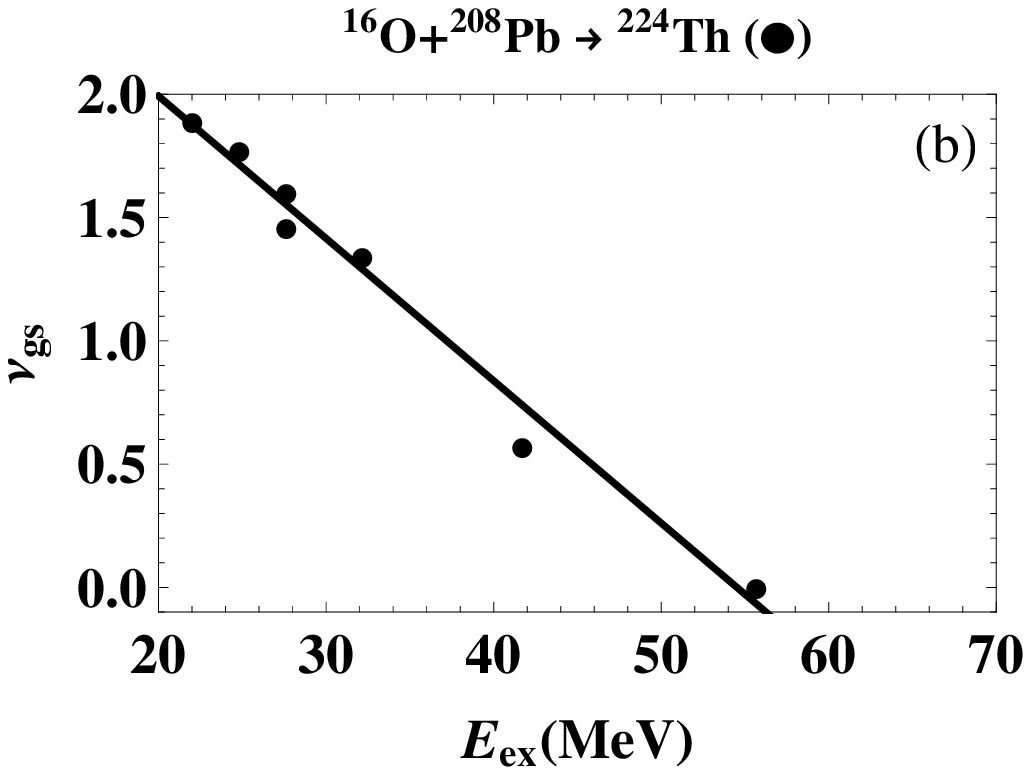}
\includegraphics[scale=0.8]{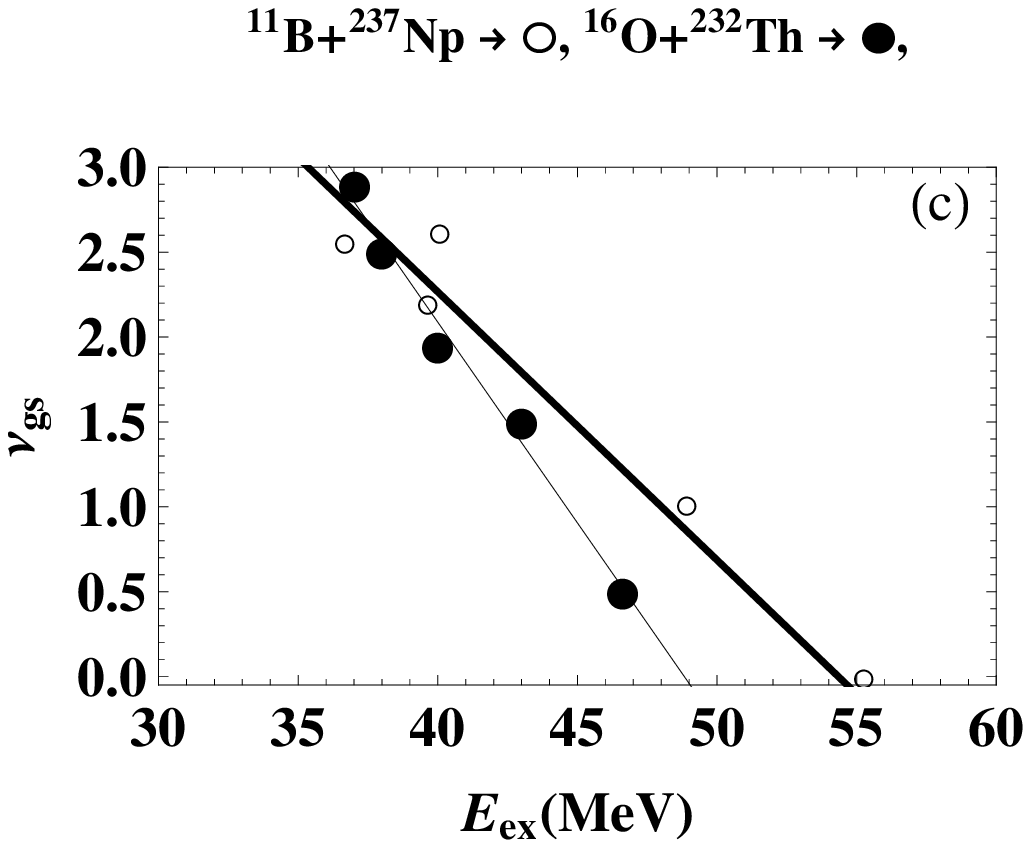}

\caption{Calculated multiplicities of pre-saddle neutrons
 (a) for the two $^{16}
\textrm{O}+^{209}\textrm{Bi}\rightarrow^{225}$ \textrm{Pa} and\\
$^{19} \textrm{F}+^{208}\textrm{Pb}\rightarrow^{227}\textrm{Pa}$
reaction systems. Thick and dotted lines represent the general
trends of $\nu_{gs}$ against the excitation energy of the compound
nucleus for the two $^{16}
\textrm{O}+^{209}\textrm{Bi}\rightarrow^{225}$ \textrm{Pa} and\\
$^{19} \textrm{F}+^{208}\textrm{Pb}\rightarrow^{227}\textrm{Pa}$
reaction systems, respectively. (b) For the
$^{16}\textrm{O}+^{208}\textrm{Pb}\rightarrow^{224}{\textrm{Th}}$
reaction system. Thick line represents the general trend of
$\nu_{gs}$ against the excitation energy of the compound nucleus,
and (c)  for the two $^{11} \textrm{B}+^{237}\textrm{Np}$ and $^{16}
\textrm{O}+^{232}\textrm{Th}$ reaction systems, both populating the
same compound nucleus $^{248}\textrm{Cf}$. Thick and thin lines
represent the general trends of $\nu_{gs}$ against the excitation
energy of the compound nucleus  for the two $^{11}
\textrm{B}+^{237}\textrm{Np}$ and $^{16}
\textrm{O}+^{232}\textrm{Th}$ reaction systems, respectively.}
\end{figure}

  The ratio of the calculated pre-saddle neutron multiplicity , $\nu^{gs}_ {cal}$
to experimental pre-scission neutron multiplicity
$\nu^{pre}_{exp}$~\cite{Rossner:1992}, and also the ratio of
theoretical pre-saddle neutron multiplicity to theoretical
pre-scission neutron multiplicity~\cite{Frobrich1: 1994}  for the
$^{16}\textrm{O}+^{208}\textrm{Pb}\rightarrow^{224}\textrm{Th}$
reaction system, are  given in Table I.
\begin{table}[h]
\begin{center}
\renewcommand{\arraystretch}{1.3}
\begin{tabular}{c  c c c c c }
\hline \hline
\textrm{$E_{c.m.}(MeV)$}~~~&~\textrm{$E_{ex}(MeV)$}~~~&~\textrm{$\nu^{pre}_{exp}$}~~~&~$\nu^{gs}_
{cal}$
&~~~~$\nu^{gs}_ {cal}/{\nu^{pre}_{exp}}$~~~~&~$\nu^{gs}_ {th}/{\nu^{pre}_ {th}}$\\
\hline
76.9&22.7& 1.50&~~ 1.81&~~ 1.21&~~ 0.96\\
82.6&27.6& 1.90&~~ 1.60&~~ 0.84&~~ 0.91\\
92.0&32.0& 2.40&~~ 1.30&~~ 0.54&~~ 0.78\\
105.9&42.5& 2.80&~~ 0.52&~~ 0.18&~~ 0.64\\
119.0&55.0& 3.40&~~ 0.00&~~ 0.00&~~ 0.56\\
\hline \hline
\end{tabular}
\caption{ Comparison between the calculated $\nu^{gs}_{cal}$,
$\nu^{gs}_{cal}/{\nu^{pre}_{exp}}$ and
$\nu^{gs}_{th}/\nu^{pre}_{th}$~\cite{Frobrich1: 1994} for the
$^{16}\textrm{O}+^{208}\textrm{Pb}\rightarrow^{224} \textrm{Th}$
reaction system.}
\end{center}
\end{table}
  As can be seen from Table I, the calculated number of pre-saddle neutrons  for  the
$^{16}\textrm{O}+^{208}\textrm{Pb}\rightarrow^{224}\textrm{Th}$
reaction system is greater than ${\nu^{pre}_{exp}}$  at
$\textrm{E}~_{\textrm{c.m.}}=76.9~ \textrm{MeV}$. This unexpected
result can be related  to the measured value  of fission fragment
angular anisotropy at low energy . It seems that the measured value
of the angular anisotropy at $\textrm{E}~_{\textrm{c.m.}}=76.9~
\textrm{MeV}$ is reported more than its actual value.

 As the nucleus is heated, the excitation  energy of the compound nucleus, $E_{ex}$
exceeds the fission barrier height, $B_f$. Hence, it becomes
possible for the nucleus to fission after passing through excited
states above the fission barrier ( transient state )~\cite{Hinde5:
1993}. In this transient state picture, the fission width,
$\Gamma_f$ depends on the level density above the fission barrier.
The fission width and the neutron width can be shown to be
approximately given by $\Gamma_f\propto\exp(-\frac{B_f}{T})$ and
$\Gamma_n\propto\exp(-\frac{B_n}T)$ ( $B_n$ is the neutron binding
energy ), respectively.  Therefore, the energy dependence of the
ratio $\frac{\Gamma_n}{\Gamma_f}$ is expected to be dominated by the
ratio of appropriate Boltzmann factors, i.e.,
$\frac{\Gamma_n}{\Gamma_f} \approx exp[(B_f-B_n)/T]$.

  In general, in heavy ion induced fission, $B_f$ will be relatively high at low excitation energy
or at low angular momentum, $I$, however as $I$, as well as $E_{ex}$
is increased, the larger moment of inertia of the elongated saddle
point configuration causes its energy to increase less rapidly than
that of the compact equilibrium deformation, so the barrier height
falls to zero at some $I$. The ratio $\frac{\Gamma_n}{\Gamma_f}$ is
known to decrease sharply as $E_{ex}$ increases in nuclei of the
Lead-Bismuth region, and it is expected to do just the opposite for
nuclei with the largest known atomic numbers~\cite{Bishop: 1972}.
For the lighter group of fissioning elements $B_f\gg{B_n}$, and for
the very heavy ones, it is expected that $B_n\gg{B_f}$. For nuclei
of intermediate mass like the Neptunium, $B_n$ and $B_f$ are nearly
equal and one expects only a slow variation of
$\frac{\Gamma_n}{\Gamma_f}$ with $E_{ex}$. In a heavy ion reaction,
there is sufficient excitation energy to emit several neutrons, and
fission can compete at each stage ( if the excitation energy is
greater than the fission barrier height ), thus the fission
probability and neutron evaporation probability at stage i, are
given by $p_{_{f, i}}= (\frac{\Gamma_{_f}}{\Gamma_{_{tot}}})_{_i}$
and $p_{_{n,
i}}=(\frac{\Gamma_{_n}}{\Gamma_{_{tot}}})_{_i}=1-(\frac{\Gamma_{_f}}{\Gamma_{_{tot}}})_{_i}$,
respectively.  As a result, the total fission probability, $P_{_f}$
is given by
\begin{equation}
P_{_f}=\sum_{k=1}^{\nu}\prod_{i=1}^{k}(p_{_{f, i}})(p_{_{n, i-1}}),
\end{equation}
where $\Gamma_{tot}=\Gamma_f+\Gamma_n$.
  The mean number of neutrons emitted before fission, $\nu_{pre}$
can be derived by the following expression
\begin{equation}
\nu_{pre}=(\frac{1}{P_f})\sum_{k=1}^{\nu}(k-1)\prod_{i=1}^{k}(p_{_{f,
i}})(p_{_{n, i-1}}).
\end{equation}
As $I$ increases, the fission barrier height decreases, then
$p_{_{f, 1}}$ along the decay chain approaches unity, and steps with
$k>1$ become insignificant, and $\nu_{pre}\longrightarrow0$; thus
fission is predicted to occur at the first step in the decay chain.
It is obvious that as the projectile energy rises, $\nu_{pre}$ will
initially rises, due to more chances for fission, but should
subsequently falls as the angular momentum reaches the value at
which $P_f$ nears  unity.  It is shown that, the transient time at
the scission point, $\tau_{sci}$ by using a diffusion model for the
fission process is given by~\cite{Hassani: 1984}
\begin{equation}
\tau_{sci}\simeq\tau_{sad}+\overline{\tau}=\beta^{-1}\ln(10B_f/T)+\overline{\tau},
\end{equation}
where, $\tau_{sad}$,  $\overline{\tau}$ and $\beta$  are the
transient time at the saddle point, the average traveling  time
between the saddle and scission points, and the nuclear friction,
respectively. The time $\overline{\tau}$ is a function of the value
of the nuclear friction, of the shape of potential and of the
excitation energy. The above equation shows that $\tau_{sad}$
depends sensibily on the nuclear friction $\beta$ and on the
excitation energy of the compound nucleus.

 Earlier calculations of fission fragment anisotropies based on SSPSM
have been corrected to include the effect of pre scission neutron
emission. The calculation of fission fragment anisotropies with
taking into account the effect of pre-scission neutron  emission
better compares  with the SSPSM predictions with the experimental
results. However, there is a small discrepancy between model
predictions and the data at high excitation energies. A fraction of
pre-scission neutrons is expected to be emitted between saddle to
scission. These latter neutrons do not longer influence the
prediction of angular anisotropy by SSPSM, since it is assumed that
the SSPSM is decided at the saddle point. In Fig. 2, the effect of
pre-saddle neutrons in the prediction of angular anisotropy by SSPSM
is demonstrated for the
$^{16}\textrm{O}+^{208}\textrm{Pb}\rightarrow^{224}\textrm{Th}$
reaction system ~\cite{Frobrich1: 1994}. As it is shown in the
figure, the discrepancy between the experimental data of angular
anisotropies and the prediction of the SSPSM can be removed to a
large extent by taking into account the pre-saddle neutron emission
correction.
\begin{figure}[h]
\includegraphics[scale=0.9]{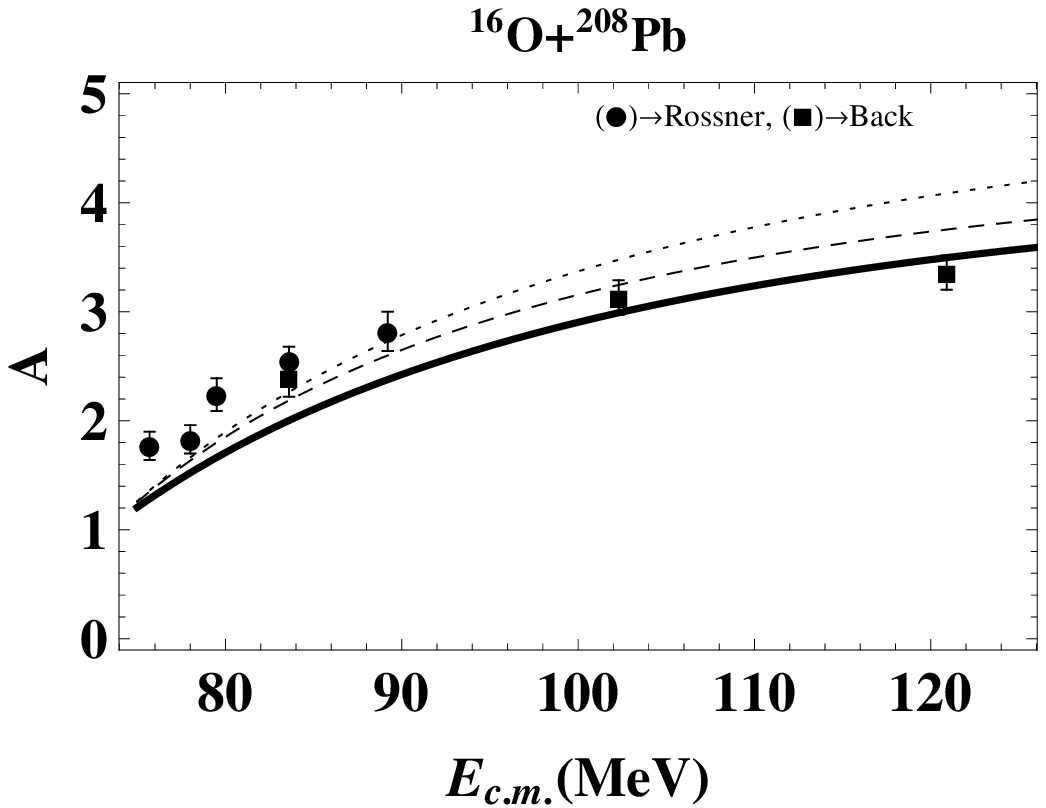}

\caption { Experimental and  calculated anisotropies in
$^{16}\textrm{O}+^{208}\textrm{Pb}\rightarrow^{224}\textrm{Th}$
reaction system [8, 54]. Thick, dashed, dotted curves are
theoretical analysis in the framework of the SSPSM without neutron
emission correction, with pre-saddle neutrons
[${\nu^{pre}_{exp.}}(\nu^{gs}_{th.}/\nu^{pre}_{th.})$] correction,
and pre-scission neutrons correction [$\nu^{pre}_{exp.}$],
respectively. }
\end{figure}
  We observe that  for  the above studied system,  the ratio of
the calculated pre-saddle neutron multiplicity to experimental
pre-scission neutron multiplicity,
$\nu^{gs}_{cal}/{\nu^{pre}_{exp}}\approx\frac{1}{4.1}$ at
$\frac{B_f}{T}=1$ is in agreement with
$\frac{\tau_{gs}}{\tau_{gs}+\tau_{ss}}\approx\frac{1}{3.7}$( where,
$\tau_{gs}$ and $\tau_{ss}$ are  ground-to-saddle and
saddle-to-scission transition times,
respectively)~\cite{Saxena1:1994}. Hence,  the neutron emission rate
by the compound nucleus in the transition from the ground state to
the saddle point and then in the transition from saddle to the
scission points are approximately uniform.

 The calculated multiplicities of pre-saddle neutrons as a function of
$E_{ex}$ for the $^{11}\textrm{B}+^{197}\textrm{Au},
^{209}\textrm{Bi}, ^{235}\textrm{U}, ^{237}\textrm{Np}$ reaction
systems are shown in Fig. 3(a). For these studied systems, the
experimental data of angular anisotropies  are taken from literature
[46, 48-50, 54-56]. The values of  $\nu_{gs}$  as a function of
$E_{ex}$ for the $^{14}\textrm{N},
^{16}\textrm{O}+^{197}\textrm{Au}$ and $^{14}\textrm{N},
^{16}\textrm{O}+ ^{209}\textrm{Bi}$ reaction systems are also shown
in Fig. 3(b). For these systems, the experimental data of angular
anisotropies are taken from literature [46, 47].  The calculated
multiplicities of pre-saddle neutrons as a function of  the
excitation energy of the compound nucleus for induced fission of the
$^{209}\textrm{Bi}$ target by using different projectiles are shown
in Fig. 3(c).

\begin{figure}[h]
\includegraphics[scale=0.8]{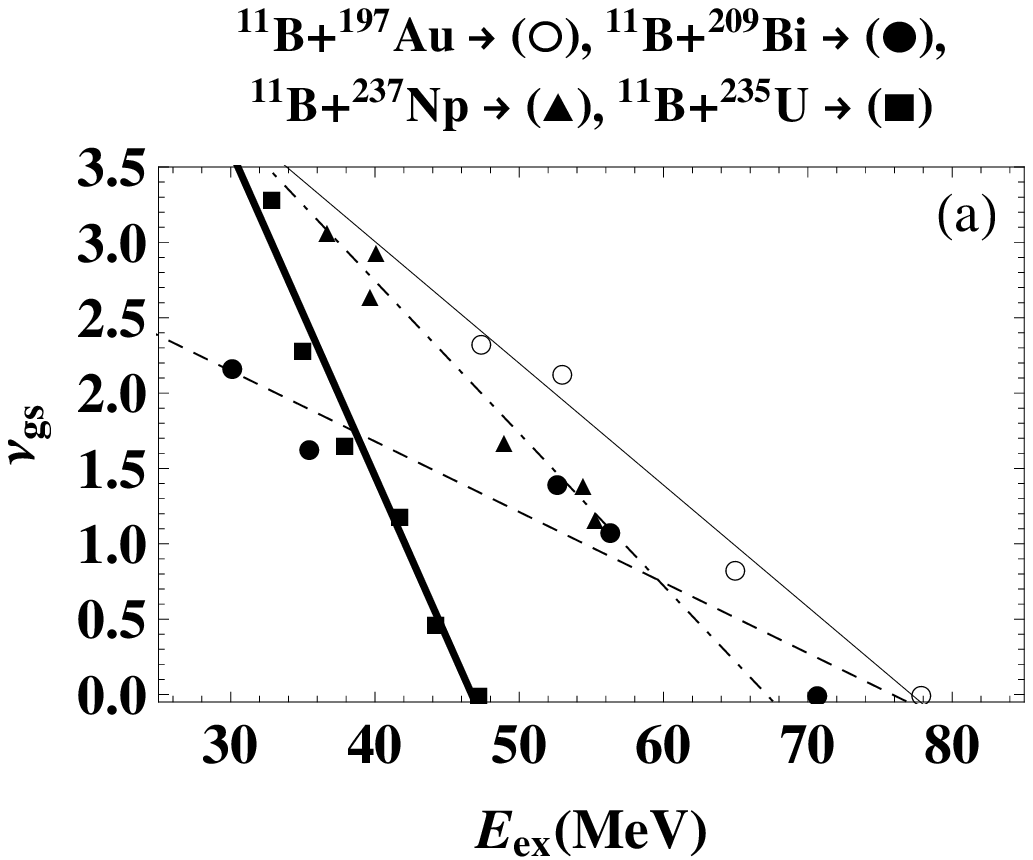}\includegraphics[scale=0.8]{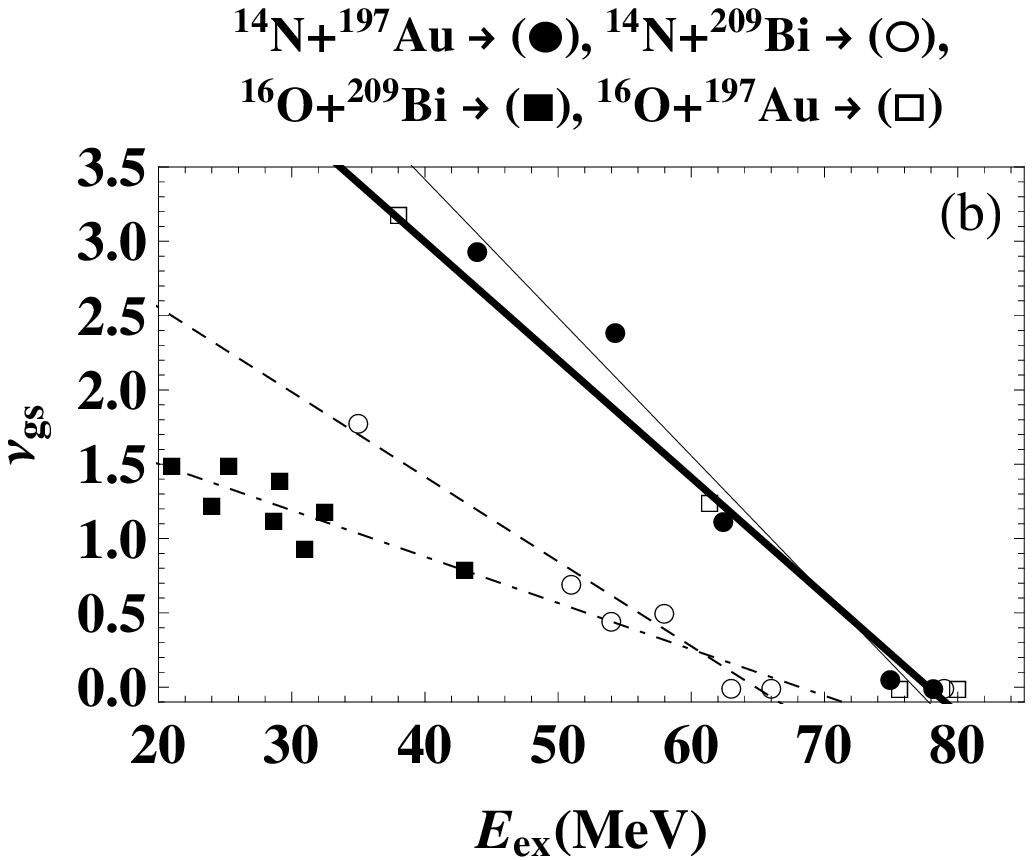}
\includegraphics[scale=0.8]{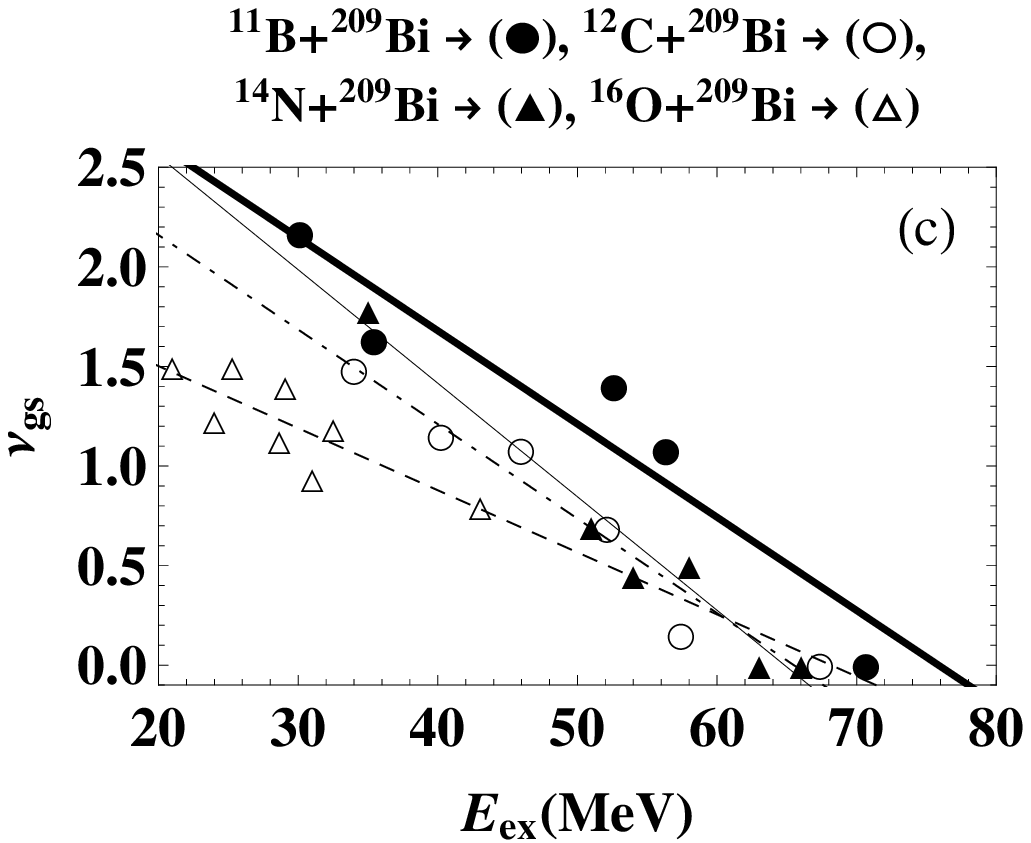}

\caption {Calculated pre-saddle neutron multiplicities, (a) for the
$^{11}\textrm{B}+^{197}\textrm{Au}, ^{209}\textrm{Bi},
 ^{235}\textrm{U}, ^{237}\textrm{Np}$ reaction systems. Thin, dashed, thick, and dashed-dotted lines
represent the general trends of the number of pre-saddle neutrons
against the  excitation energy of the compound nucleus,
respectively. (b) for the $ ^{14}\textrm{N},
^{16}\textrm{O}+^{197}\textrm{Au}$ and  $^{14}\textrm{N},
 ^{16}\textrm{O}+ ^{209}\textrm{Bi}$ reaction systems. Thin,  thick, dashed, and dashed-dotted lines
represent the general trends of the number of pre-saddle neutrons
against the excitation energy of the compound nucleus for these
systems, respectively., and (c)  for the $ ^{11}\textrm{B},
^{12}\textrm{C},^{14}\textrm{N}, ^{16}\textrm{O} +^{209}\textrm{Bi}$
reaction systems. Thick, dashed-dotted, thin, and dashed lines
represent the general trends of the number of pre-saddle neutrons
against the excitation energy of the compound nucleus for these
systems, respectively.
   }
\end{figure}

 The average values of $\nu_{gs}$, as well as ranges of  pre-saddle neutron multiplicities
for the fission reactions of different targets induced by the same
projectile over the same projectile energy range are shown in Table
II. In this Table, the quantity $V_{b}$ denotes the Coulomb barrier
height. It can be observed that $\overline{\nu}_{gs}$ decreases with
increasing the  mass number of the target.

\begin{table}[f]
\begin{center}
\renewcommand{\arraystretch}{1.3}
\begin{tabular}{c c c c}
\hline \hline
\textrm{reaction systems}&~~~\textrm{Projectile energy
(in $\frac{E_{c.m.}}{V_{b}}$)}&$\nu_{gs}$&$\quad$\textrm{$\overline{\nu}_{gs}$}\\

\hline
$^{11}\textrm{B}+^{197}\textrm{Au}$&1.4-1.9&3.1-1.4& 2.0\\
$^{11}\textrm{B}+^{209}\textrm{Bi}$&1.4-1.9&1.8-0.8& 1.6\\
$------$&$------$&$------$&$~~-------$\\
$^{12}\textrm{C}+^{197}\textrm{Au}$&1.3-1.8&2.4-1.6& 2.1\\
$^{12}\textrm{C}+^{209}\textrm{Bi}$&1.3-1.8&1.5-0.4& 1.0\\
$------$&$------$&$------$&$~~-------$\\
$^{14}\textrm{N}+^{197}\textrm{Au}$&1.2-1.7&3.0-0.5& 1.9\\
$^{14}\textrm{N}+^{209}\textrm{Bi}$&1.2-1.7&1.6-0.1& 0.9\\
$------$&$------$&$------$&$~~-------$\\
$^{16}\textrm{O}+^{197}\textrm{Au}$&1.0-1.6&3.3-0.7& 2.0\\
$^{16}\textrm{O}+^{208}\textrm{Pb}$&1.0-1.6& 1.9-0.1&1.5\\
$^{16}\textrm{O}+^{209}\textrm{Bi}$&1.0-1.6&1.7-0.9& 1.4\\
 \hline \hline
\end{tabular}
\caption{ Comparison between the calculated pre-saddle neutron
multiplicity in the form of a range, as well as
$\overline{\nu}_{gs}$ for  fission reactions of the different
targets induced by the same projectile.}
\end{center}
\end{table}
 The average values of $\nu_{gs}$, as well as the pre-saddle neutron multiplicity in the form of
a range for the induced fission of the same target by different
projectiles over the same projectile energy are also given in Table
III. As can be seen  in Table III, the quantity
$\overline{\nu}_{gs}$ decreases with increasing the  mass number of
projectile. All heavy ion induced reactions show that $\nu_{gs}$
falls quite rapidly with increasing the mass asymmetry, since it is
partly due to a reduction of the dynamical fission time scale with
the mass asymmetry.

\begin{table}[f]
\begin{center}
\renewcommand{\arraystretch}{1.3}
\begin{tabular}{c c c c}
\hline \hline \textrm{reaction systems}&~~~\textrm{Projectile energy
(in $\frac{E_{c.m.}}{V_{b}}$)}&$\nu_{gs}$&$\quad$\textrm{$\overline{\nu}_{gs}$}\\

\hline
$^{11}\textrm{B}+^{209}\textrm{Bi}$&1.2-1.7&2.2-1.2&1.9\\
$^{12}\textrm{C}+^{209}\textrm{Bi}$&1.2-1.7&1.8-0.8& 1.2\\
$^{14}\textrm{N}+^{209}\textrm{Bi}$&1.2-1.7&1.6-0.2&0.9\\
$^{16}\textrm{O}+^{209}\textrm{Bi}$&1.2-1.7&1.4-0.6&0.8\\
$------$&$------$&$------$&$~~-------$\\
$^{12}\textrm{C}+^{197}\textrm{Au}$&1.2-1.6&2.8-1.8& 2.4\\
$^{14}\textrm{N}+^{197}\textrm{Au}$&1.2-1.6&3.0-1.0& 2.3\\
$^{16}\textrm{O}+^{197}\textrm{Au}$&1.2-1.6&2.5-0.7& 2.0\\
$------$&$------$&$------$&$~~-------$\\
$^{16}\textrm{O}+^{208}\textrm{Pb}$&1.1-1.6&1.9-1.0& 1.9\\
$^{19}\textrm{F}+^{208}\textrm{Pb}$&1.1-1.6&1.4-0.4& 1.4\\
 \hline \hline
\end{tabular}
\caption{ Comparison between the calculated pre-saddle neutron
multiplicity in the form of a range, as well as
$\overline{\nu}_{gs}$ for  fission reactions of the same target
induced
 by different heavy ions.}
\end{center}
\end{table}

  We now attempt to estimate the pre-saddle neutron multiplicities
in several fission reactions  induced by light projectiles. We must
pay attention to some important points expressing the difference
between  fission induced by light projectiles and heavy ions. In the
fission induced by light projectiles, the energy in the
center-of-mass framework, $E_{c.m.}$ is roughly the same as that in
the laboratory framework, as well as due to the low weight of
projectile, rotational energy, $E_{R}$ can be neglected. Fig. 4,
shows calculated pre-saddle neutron multiplicities for the two
$\alpha+^{182}\textrm{W}$, and $\textrm{p}+^{185}\textrm{Re}$
reaction systems which are leading to similar $^{186}\textrm{Os}$
compound nucleus, as well as for the two
$\textrm{p}+^{209}\textrm{Bi}$, and $\alpha+^{206}\textrm{Pb}$ that
formed the same  $^{210}\textrm{Po}$ compound nucleus. For these
systems, the experimental data of angular anisotropies are taken
from literature [57-59]. The values of $<I^{2}>$  for these systems
are given by~\cite{Ignatyuk1: 1984}:
\begin{equation}
<I^{2}>=\frac{\sum(2I+1)T_{I}I(I+1)}{\sum(2I+1)T_{I}}
\end{equation}
where $T_{I}$ is the entrance channel transmission coefficients and
satisfy $T_{I}=1$ for $I\leq{I_{max}}$ and $T_{I}=0$ for
$I>I_{max}$. If the maximum angular momentum is determined by the
relation $<I^{2}>=1/2I_{max}^{2}$, the following relations give the
values of the mean square angular momentum of the compound nucleus
for the  fission of pre-actinide nuclei induced by proton and
$\alpha$ particle, respectively:
\begin{equation}
<I^{2}>=2.08E_{p}(MeV)-15,
\end{equation}
\begin{equation}
<I^{2}>=10.2E_{\alpha}(MeV)-199.
\end{equation}

 In heavy ion induced fission at low bombarding energies, several
neutrons are evaporated prior to the reaching to the saddle point,
and at the highest bombarding energy essentially all the neutrons
are evaporated by the fission fragments, i.e., the fission process
is rapid compared to the time scale for neutron evaporation.
However, the number of pre-saddle neutrons, $\nu_{gs}$ increases
with increasing the excitation energy of the compound nucleus in
fission induced by light projectiles. This behavior is mainly due to
that in the induced fission by light projectile, the fission barrier
height is higher than the neutron binding energy, as well as $B_f$
is approximately independent of the excitation energy of the
compound nucleus. Therefore, the fission probability,
$P_f=\frac{\Gamma_f}{\Gamma_{tot}}$ is negligible at low energies.
When $E_{ex}<{B_{f}}$, it is impossible that the compound nucleus
undergoes fission, but there is sufficient excitation energy to emit
several neutrons. It is clear that the fission becomes significant
if $E_{ex}>{B_{f}}$.
\begin{figure}
\includegraphics[scale=0.9]{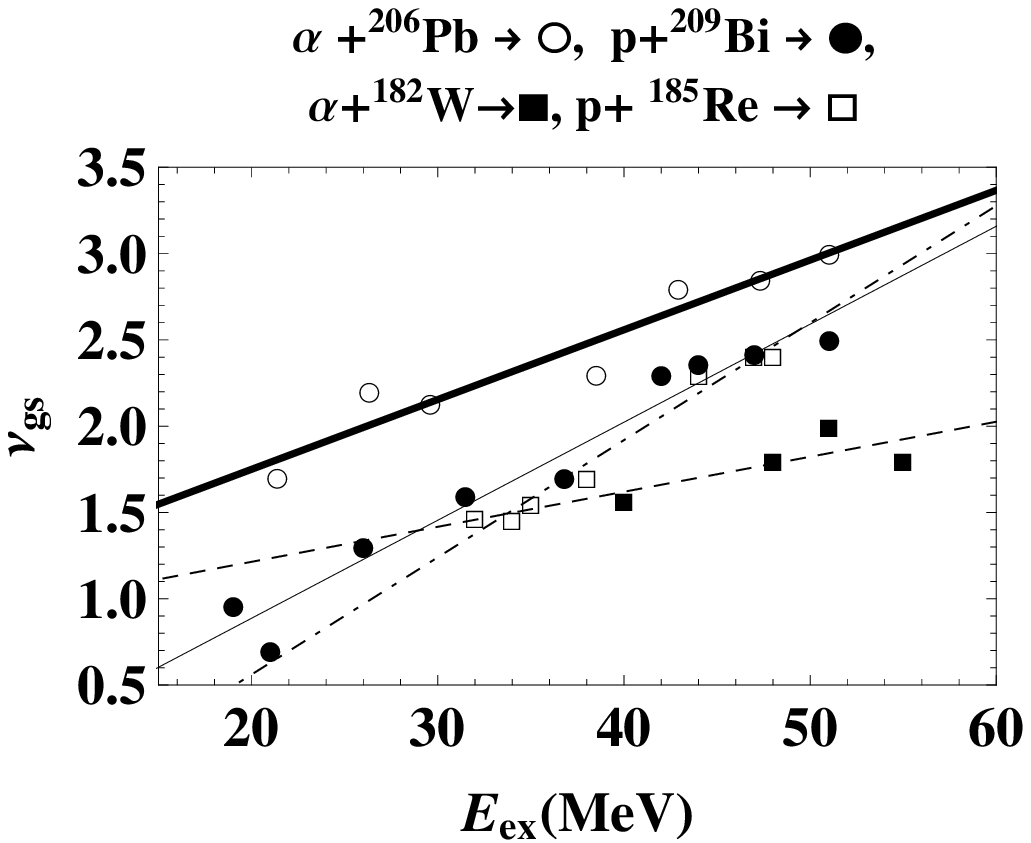}

\caption{The values of $\nu_{gs}$ for the two
$\alpha+^{206}\textrm{Pb}$ , and $\textrm{p}+^{209}\textrm{Bi}$
reaction systems which are leading to the similar
$^{210}\textrm{Po}$ compound nucleus, as well as for the two
$\alpha+^{182}\textrm{W}$, and $\textrm{p}+^{185}\textrm{Re}$
reaction systems that formed the same $^{186}\textrm{Os}$ compound
nucleus. Thick, thin, dashed, and dashed-dotted lines represent the
general trends of the pre-saddle neutrons against the excitation
energy of the compound nucleus for these systems, respectively.}
\end{figure}

\section{Summary and Conclusions}
We have presented in this paper the calculated pre-saddle neutron
multiplicities  for several heavy ion induced fission reactions, as
well as for several  fission reactions  induced by light
projectiles. The calculation  by using the experimental data of
fission fragment angular anisotropies, as well as the prediction of
the SSPSM is a novel method, which has been carried out in this work
for the first time. We have also considered the behavior of
pre-saddle neutron multiplicities in  fission reactions induced by
heavy ions and light projectiles. In heavy ion induced fission, the
number of pre-saddle neutrons decreases with increasing the
excitation energy of the compound nucleus. Whereas in fission
induced by light particles, the number of pre-saddle neutrons
increases with increasing the excitation energy of the compound
nucleus. The fission barrier height in heavy ion fission reaction
depends on the excitation energy of the compound nucleus. On the
other hand, the fission barrier height ( and thus ground-to-saddle
transition time ; $\tau_{gs}\propto\ln(10B_f/T)$ ) decreases with
increasing the excitation energy of the compound nucleus. As a
result, in heavy ion induced fission the number of pre-saddle
neutrons decreases with increasing the excitation energy of the
compound nucleus. Our results also shows that the emission rate of
neutrons is approximately  constant in transition from the ground
state  to the saddle point and then from the saddle to the scission
points. On the contrary, in  fission induced by light projectiles,
the fission barrier height is greater than the neutron binding
energy, and  the fission barrier is approximately independent of the
excitation energy.  Hence, the compound nucleus does not undergo
fission, unless the excitation energy of the compound nucleus
exceeds the fission barrier. As a result,   in  fission induced  by
light projectiles, the number of pre-saddle neutrons exhibits an
increasing function against the excitation energy of the compound
nucleus as shown our calculations. The number of pre-saddle neutrons
for reactions lead to the same compound nucleus at any given
excitation energy appears to be nearly equal, since the number of
pre-saddle neutrons depends only on the mass of the compound nucleus
and it is independent of the entrance mass asymmetry parameter. This
behavior of the number of the pre-saddle neutrons as a function of
the projectile mass and/or of the target mass may also be related to
the size of compound nucleus. We observe that the average number  of
pre-saddle neutrons decreases with increasing the mass number of
projectile in fission reactions of the same target induced by
different projectiles. A similar behavior in the multiplicities of
pre-saddle neutrons  is also observed in fission reactions  of
different targets induced by the same projectile. At the end, our
results may provide useful information on the  ground
state-to-saddle and saddle-to-scission transition times.

\end{document}